\documentclass[11pt]{article}

\usepackage{geometry}
\usepackage{fourier}
\usepackage[T1]{fontenc}
\usepackage{paralist}
\usepackage{amssymb}
\usepackage{graphicx}
\usepackage{microtype}
\usepackage[svgnames]{xcolor}
\usepackage{xspace}
\usepackage{multicol}

\usepackage[hyphens]{url}
\usepackage{hyperref}
\definecolor{mylinkcolor}{RGB}{0,0,160}
\hypersetup{colorlinks,allcolors=mylinkcolor,citecolor=mylinkcolor}
\usepackage[capitalize,noabbrev]{cleveref}

\usepackage{doi}

\let\OLDthebibliography\thebibliography
\renewcommand\thebibliography[1]{
  \OLDthebibliography{#1}%
  \footnotesize%
  \setlength{\parskip}{0pt}%
  \setlength{\itemsep}{0pt plus 0.3ex}%
}

\newcommand{\emhdr}[1]{\medskip\noindent{\textbf{\emph{#1.}}}}

\title{\vspace*{-2em}
Higher-order Network Analysis Takes Off, \\ Fueled by Classical Ideas and New Data%
\footnote{This article is based on one published online by SIAM News~\cite{higher-order-sinews}.}}
\author{%
\small{Austin R.~Benson, Cornell University (arb@cs.cornell.edu)} \\
\small{David F.~Gleich, Purdue University (dgleich@purdue.edu)} \\
\small{Desmond J.~Higham, The University of Edinburgh (d.j.higham@ed.ac.uk)}
}

\date{}

\begin{document}
\maketitle

\vspace*{-5mm}
\begin{abstract}
Higher-order network analysis uses the ideas of hypergraphs, simplicial
complexes, multilinear and tensor algebra, and more, to study complex
systems. These are by now well established mathematical abstractions. What's new
is that the ideas can be tested and refined on the type of large-scale data
arising in today's digital world.
This research area therefore is making an impact across many applications.
Here, we provide a brief history, guide, and survey.
\end{abstract}
\vspace*{1mm}

\section*{Higher-order Network Analysis}

In 2004, the theme of Mathematics Awareness Month was ``the mathematics of networks.''\footnote{\url{https://web.archive.org/web/20041204022508/http://mathforum.org/mam/}}
A corresponding SIAM News article~\cite{yearofthenetwork} dubbed 2004 ``The Year of the Network'' and predicted that graphs would soon be everywhere.
Now, networks and graphs were not new in 2004. 
The first use of graphs in mathematics dates back to the 1800s and ``chemicographs''~\cite{Sylvester1878},
and the origins of PageRank-style linear algebra can be traced back to the same century in the context of chess player ranking~\cite{Landau95}.
Graph algorithms were common in the 1950s and served as a fascinating and fertile area in the new discipline of computer science. 
By 2004, their time had arrived. 
New systems with network and graph structures, notably from biological and internet settings, suggested novel questions. 
We believe Fan Chung Graham's observation at the time~\cite{graham2004large} was prescient:
\begin{quote}
\emph{In similar ways, many of the information networks that surround us today provide interesting motivation and suggest new and challenging research directions that will engage researchers for years to come.}
\end{quote}

Since 2004, the opportunities in what we would now call network science and data science grew tremendously due to the recognition of the power of computational techniques to tease subtle insights out of networked data. This led to more data, and more questions, as Fan Chung Graham predicted. Yet some of those questions could not easily be addressed with the standard representations of a complex system as a graph. 

To appreciate this point, it is useful to make a distinction 
between the terms network and graph. One framing that captures the essential difference is:
\begin{quote}
\emph{a network is a graph that represents a complex system.}
\end{quote}
So all networks are graphs but there are graphs---say the Petersen graph, which was constructed as a counterexample to a mathematical conjecture---that are not networks. A graph consists of vertices and edges, and each edge is comprised of a pair of vertices. 
Network analysis and network science is the field that grew up around using the formalism of graphs to simplify and study more complicated systems from the internet, social science, biology, neuroscience, and many other applications. 

\begin{figure}[t]
\centering
\includegraphics[width=0.80\columnwidth]{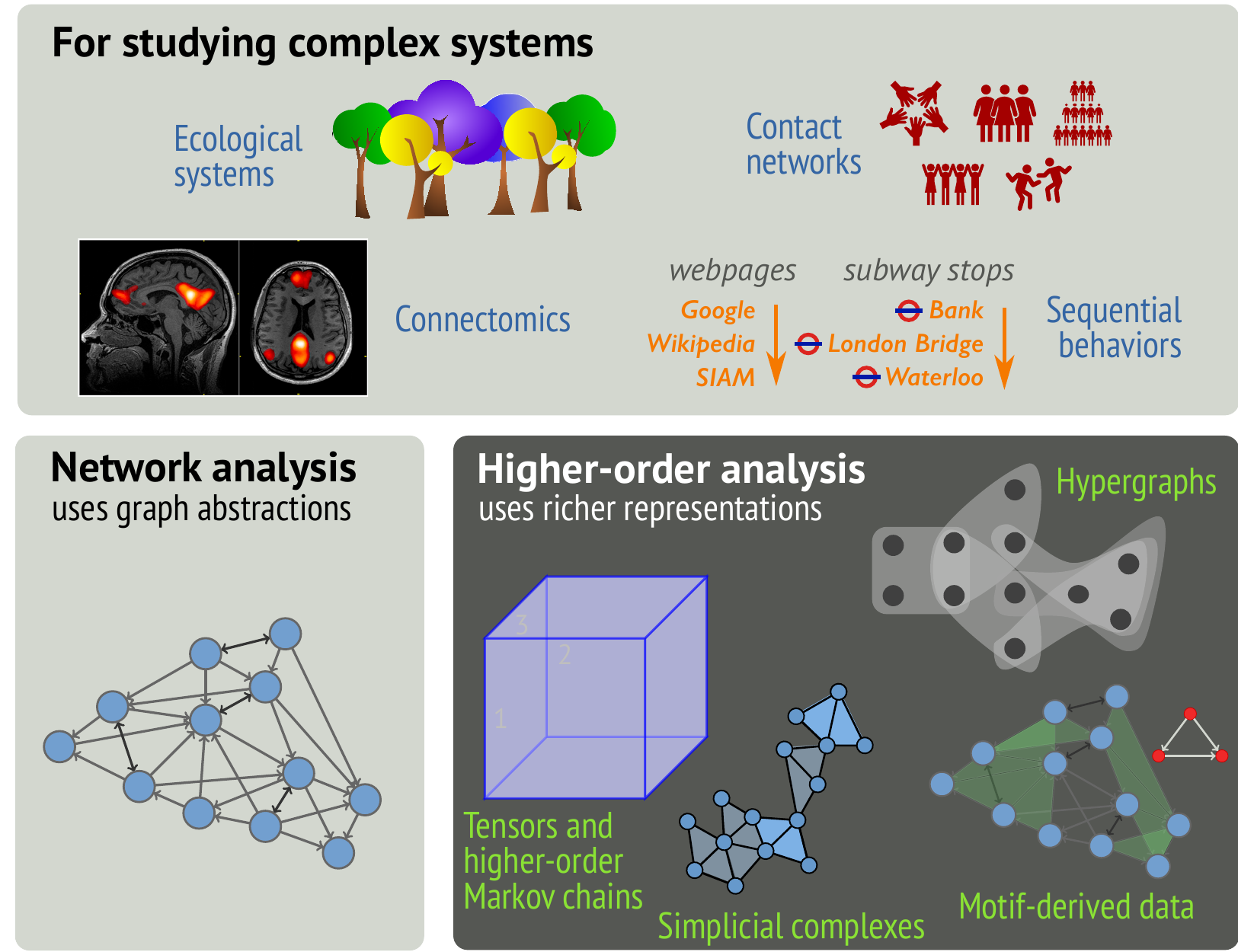}
\caption{Higher-order network analysis involves richer representations of complex systems to gain deeper insight into data. Tree figure from~\cite{grilli2017higher}; connectome image from neurosynth website~\cite{Yarkoni-2011-neurosynth}. }
\label{fig:ho_fig}
\end{figure}

Crucial to our story is that this framework offers the opportunities of richer representations of complex systems (\cref{fig:ho_fig}). 
For instance, the community quite rapidly pointed out that 
the pairwise relationships modeled by edges were insufficient to understand many properties. Milo et al.'s trendsetting work on motifs in biological signalling systems~\cite{Milo-2002-motifs} identified and attempted to interpret small circuits that appear as statistically significant substructures. 
Likewise, triangles, transitivity, and triadic completion play a key role in understanding social network data---as pointed out by  
Rapoport~\cite{rapoport1953spread},
Davis~\cite{davis1970clustering}, and
Granovetter~\cite{granovetter1973strength}
during the 1950s through the 1970s in research that predates the rapid growth of social network analysis.

The idea of multinode interactions is also not entirely new. 
Hypergraphs, those where edges represent a multiway connection, were discussed in the 1960s and 1970s, in
for example, Berge's book \emph{Graphs and Hypergraphs}~\cite{berge-book}. 
These proved to be important in the layout of VLSI circuits~\cite{hmetis-vlsi} and partitioning computations for the emerging class of parallel distributed computers based on MPI~\cite{hendrickson2000-graph-partitioning}.

It is also important to appreciate that a graph is a wonderfully flexible concept. 
A weighted graph may simultaneously represent a Markov chain, a sparse matrix, a diffusion operator, and many other objects,  depending on its exact use. Even problems posed on hypergraphs may reduce to a graph computation;
for instance, one can find the connected components of a hypergraph by computing connected components of a related bipartite graph.

But not all problems reduce to graphs. 
For example, when Rosvall et al.~wanted to understand traffic patterns in transportation systems such as subways and airplanes, it became clear that traditional memoryless Markov chains were inappropriate~\cite{Rosvall-2014-memory}. A similar scenario played out when scientists at Yahoo!\ wanted to understand web browsing behavior~\cite{Chierichetti-2012-web-users} and found a straightforward Markov chain too crude a tool. To address issues 
arising from the data, each reached for new types of higher-order stochastic processes with more memory, namely higher-order and variable-order Markov chains.

As the number of these scenarios and questions grew, a multitude of rich and higher-order representations emerged. These include tensors, simplicial complexes, and hypergraphs.
Such variations arise because the appropriate representation depends crucially on the research problem being addressed. 
Here is an illustration around disease propagation,
which is adapted from an example given by Torres et al.~\cite{torres2020representations}:
\begin{description}
\item[1.] \emph{Has this pair of individuals ever been close enough to infect each other?} \\
In this case, a \textbf{simple undirected graph} is appropriate.
This is the traditional setting where each edge records a possible route for infection.

\item[2.] \emph{Has this set of individuals ever formed all or part of a group that came close enough to infect each other?} \\
In this case, a \textbf{simplicial complex} is appropriate.
This structure allows for downward closure: any subset of nodes within a simplex
also forms a simplex.
Each simplex identifies a set of people who might be infected simultaneously.

\item[3.] \emph{Has this set of individuals ever made up an entire group that came close enough to infect each other?} \\
In this case, a \textbf{hypergraph} is appropriate.
A proper subset of those individuals will not appear as a hyperedge unless they have gathered as a complete group.
This is perhaps the most natural and informative structure, where we record only entire groups of individuals who might be infected simultaneously.
\end{description}

These new modeling methodologies can also feature the behavior of \emph{processes}, such as random walks, on top of the data. 
Consider this example~\cite{aksoy2020hypernetwork,lu2011high},
 where we are concerned with the behavior of a random walk-like process on a hypergraph:
\begin{enumerate}
    \item[1.] Suppose we consider a random walk that visits the nodes of a hypergraph by picking a random hyperedge that involves the current node and then picks a random node in that hyperedge. 
    This process is equivalent to a random walk on the bipartite network that represents the hypergraph where we ``ignore'' or ``censor'' the nodes representing the hyperedges.
    This, in turn, corresponds to a random walk on a weighted graph with the same node set as the hypergraph, and
    the process is Markovian on the state space of the nodes.
    In this case, the questions about the process can then be answered via existing graph-based and matrix-based techniques.
  \item[2.] Consider the following equivalent description of that process to motivate our generalization. Given a starting node, the process defines a sequence of hyperedges where adjacent edges in the sequence share a single node.  Suppose, however, we make a tiny modification so that adjacent edges must share \emph{two nodes}. This new procedure defines an edge process that cannot easily be represented using traditional graph and matrix techniques, as there is no notion of ``where'' the walk is that admits a simplification. 
  The process is no longer Markovian on the state space of the nodes but more complex machinery will compute probabilities associated with the process via a related Markov chain.
\end{enumerate}
Thus, new developments in higher-order analysis begin to appear when the studies are more specialized and focused. 

\section*{Higher-order machinery}

Although higher-order network analysis is still being actively researched and developed, there are a number of 
technical tools and ideas that are becoming common. We'll 
summarize some key examples and see how they work out in terms of generalizations of PageRank. 

\emhdr{Hypergraphs} A hypergraph is simply a generalization of a graph structure where edges can represent relationships among more than two entities. They are perhaps the most general representation of higher-order relationships. Mathematical functions can be associated with each hyperedge to make the setting extremely general. 

\emhdr{Motifs} Motifs were among the first formalizations to accommodate the idea that small patterns could be more important than just edges. Finding all instances of a motif or set of motifs, then, serves as a way of extracting or finding higher-order structure within simple graph data. It can also be a way of simplifying a collection of complex data into a more concretely specified problem. A collection of motifs of the same size may be stored as a uniform hypergraph (where all hyperedges have the same size).

\emhdr{Higher-order Markov chains}
Markov chains and random walks are canonical stochastic processes on a graph where the behavior is independent of the past behavior of the process. Generalizations are therefore natural. Higher-order Markov chains are stochastic processes that depend on a limited space of past behavior. Many computations on these reduce to a standard Markov chain on a Cartesian product state-space. However, the mathematically interesting relationships start to happen when considering approximations and variable-order chains. Non-backtracking
random walks~\cite{krzakala2013spectral} (and their generalizations~\cite{RStensor,arrigo2019nonbacktracking})---which are higher-order Markov chains where the process does not immediately return to its previous state---are now considered to be a standard analysis tool, which raises new questions and algorithmic possibilities.

\emhdr{Tensors and hypermatrices}
A tensor or hypermatrix is a multidimensional generalization of a matrix. Rectangular tensors have long been studied in terms of low-parameter structure~\cite{harshman1970foundations}.
The adjacency tensor of a $k$-element uniform hypergraph is an equal-sided (or cubical) array where all entries associated with a hyperedge (and all the permutations!) have value $1$ and all other entries have value $0$. Tensor eigenpairs arise in generalizations of graph centrality measures to uniform hypergraphs~\cite{arrigo2020framework,benson2019three} and in generalizations of PageRank, as we'll explain shortly. 
One downside to tensors shown by Hillar and Lim~\cite{hillar2013most} is that most tensor problems are NP-hard. 
This motivates work to identify structures in the tensors that enable polynomial time algorithms, such as spacey random walks~\cite{benson2017spacey} and new types of ergodicity coefficients~\cite{fasino2020ergodicity}.

\emhdr{Simplicial complexes}
A simplicial complex is a collection of sets (simplices) that is closed with respect to taking subsets---for any $X$ in the complex, any nonempty subset $Y$ of $X$ is also in the complex.
A graph can be thought of as a simplicial complex, where each edge and vertex is mapped to a set,
or richer graph structure can be encoded in a simplicial complex by mapping each clique of size at most $k$ to a set.
Similarly, one can induce a simplicial complex from a hypergraph by treating the edges as sets of nodes and including all of the edges and their subsets.
Modeling interactions with simplicial complexes naturally leads to using tools from topology.
For example, homology, cohomology, and Hodge theory have been used to
find topological holes or cavities important in brain networks~\cite{sizemore2018cliques},
identify voting ``islands'' within elections~\cite{feng2021persistent}, and
estimate traffic in road systems~\cite{jia2019graph,roddenberry2019hodgenet}.
Matrix computations involving generalizations of the graph Laplacian to Hodge Laplacians often underlie these methods~\cite{lim2020hodge}.

\subsection*{Examples of these tools with PageRank}

PageRank~\cite{page1999-pagerank} is a central algorithm in network science~\cite{gleich2015pagerank}. 
The elegance of the mathematics behind PageRank enables many possible derivations.
The classic derivation is via Markov chains and random walks and the random surfer on the web.
Alternative derivations arise via
social theory~\cite{friedkin1991theoretical},
smooth functions on graphs~\cite{zhou2003learning},
biased and localized spectral clustering~\cite{mahoney2012local},
inference of blockmodels~\cite{kloumann2017block},
mincut minorants~\cite{gleich2015using}, and
edge-based walks~\cite{arrigo2019non}.
These various derivations, in appropriate regimes, all correspond to the same mathematical expression involving a linear system on a graph. 

And there are higher-order generalizations of many of these ideas.
However, the difference is that these higher-order generalizations (typically) fail to have any relationship.
Some are nonlinear.
Some are extremely specific to the data representation.

One can, for instance, take any higher-order Markov chain and study a random surfer.
This produces a related higher-order Markov chain.
Using a spacey random walk on that same higher-order Markov chain, however, yields a tensor eigenvector~\cite{benson2017spacey,gleich2015multilinear}.
There are related notions of PageRank on hypergraphs~\cite{li2020quadratic,liu2021strongly} and motif-derived hypergraphs~\cite{yin2017local}. 
These use the ability of hypergraph edges to be associated with functions that model complex cuts~\cite{li2018submodular,veldt2020hypergraph}.
There is also PageRank on a simplicial complex, which uses random walks derived from normalizing a Hodge Laplacian matrix, 
just as one can derive PageRank by normalizing a graph Laplacian matrix to get a random walk matrix~\cite{schaub2020random}.

Each generalization has different modeling and computational tradeoffs. 
PageRank as a tensor eigenvector does not have a clear algorithmic standing and we have conjectured it is PPAD-complete. 
Hypergraph PageRank is a convex problem when using the Lov\'{a}sz extension of a submodular function associated with the hyperedge. 
For $3$-uniform hyperedges, this reduces to a simple matrix computation. 
The simplicial complex PageRank vector has an orthogonal decomposition into three components with different topological meanings.

Which of these is the most appropriate depends on the nature of the question. 
While higher-order tools enable richer and more nuanced insights into data, they also demand a keen appreciation of the implicit assumptions and applied semantics of the methods, as illustrated in the contact-infection example above. 

\section*{Impacts and Open Possibilities}

In additional to fascinating mathematical and computational issues, higher-order methods are of growing importance in many studies. 

\begin{compactenum}
    \item Research into ecology systems showed that multiway interactions among species are needed to resolve paradoxes between real-world ecological outcomes and simple synthetic pairwise competition models~\cite{grilli2017higher}. 
    \item Within studies of neuroscience, higher-order methods are conjectured to be essential ``in unraveling the fundamental mysteries of cognition''~\cite{giusti2016two}.
    \item Accurate modeling and combinatorics for certain types of traversal can account for higher-order structure,
     in applications ranging from 
    ecosystems \cite{ER06} to passenger air travel~\cite{Rosvall-2014-memory}.
    Such flows can even be used to define what it means to be ``higher-order''~\cite{Lambiotte-2019-higher-order}. 
  \item For dynamical systems more broadly, higher-order interaction models result in markedly new behavior in many settings~\cite{battiston2020networks},
    such as opinion dynamics~\cite{neuhauser2020multibody}, synchronization~\cite{Carletti2020}, and disease spread~\cite{IPBL19}.
\end{compactenum}

There remain numerous interesting questions around these methods. For example, there appears to be something special that goes on for three-node patterns that simplifies models on hypergraphs in terms of cuts down to a simple matrix case. 
Showing that this is true is relatively straightforward~\cite{benson2016higher,li2020quadratic}, but we presently lack a clear appreciation of \emph{if} this is a simple artifact of the mathematics \emph{or} if there might be some more general theory that could be deployed in new scenarios. 
Many of the most important challenges in network science and higher-order network analysis are matters of statistical or causal relevance.
These are not yet standardized and rigorous methods are 
required to assess findings of significance in networks---although there is excellent research in the community on these issues that offers a number of ways forward. 
Finally, there is a pressing need for efficient and rigorously justified algorithms
that can accommodate and enable the possibilities of higher-order analysis. This includes the overriding challenge of optimizing computation of higher-order quantities as problems get larger. 

\bigskip\bigskip
\noindent \textbf{Acknowledgments.}
\footnotesize
This work was supported in part by
ARO W911NF19-1-0057, 
ARO MURI, 
NSF DMS-1830274, 
JP Morgan Chase \& Co, 
NSF IIS-1546488, 
NSF CCF-1909528, 
NSF IIS-2007481, 
the NSF Center for Science of Information STC, CCF-0939370, 
NASA, 
Sloan Foundation, and
EPSRC Programme Grant EP/P020720/1.

\begin{multicols}{2}
\bibliographystyle{plainurl}
\bibliography{refs}
\end{multicols}
\end{document}